\documentclass[pra,aps,showpacs,amssymb]{revtex4}
\usepackage{epsf,psfig,graphicx,latexsym,amsmath}
\usepackage{supercite}
\def\comment#1{}

\newcommand{\BF}[1]{\mbox{\boldmath $#1$}}
\begin{document}
\title{ Quantum Phase Diagram\\ for Homogeneous Bose-Einstein Condensate}
\author{Hagen Kleinert$^{1}$, Sebastian Schmidt$^{1,2}$, and Axel Pelster$^{1}$}
\affiliation{$^{1}$Institut f\"ur Theoretische Physik,
Freie Universit\"at Berlin, Arnimallee 14, 14195 Berlin, Germany\\
$^{2}$Department of Physics, Yale University, P.O. Box 208120, New Haven, CT 06520-8120, USA}
\begin{abstract}
We calculate
the quantum phase transition for
a  homogeneous Bose gas in the plane
of $s$-wave scattering length $a_s$ and temperature $T$.
This is done by improving a
one-loop result near
the interaction-free Bose-Einstein critical
temperature $T_c^{(0)}$
with the help of
recent high-loop results on the shift of the critical
temperature due to a weak atomic repulsion
using variational perturbation theory.
The quantum phase diagram shows a {\it nose} above
 $T_c^{(0)}$, so that we predict the existence of a
reentrant transition {\it above}   $T_c^{(0)}$, where
an
{\it increasing} repulsion
leads to the formation
of a condensate.
\end{abstract}
\maketitle
\section{Introduction}
So far, experimental work on Bose-Einstein condensates
(BECs) has dealt with magnetic traps, so that
most of the recent theoretical investigations focus on such
systems
\cite{Pitaevskii,Leggett,Pethick,Stringari}.
However, there are interesting
 theoretical problems also in homogeneous BECs
with weak two-particle interactions.
For instance, a very fundamental property of homogeneous BECs
with a local repulsive interaction
$V^{({\rm int})}({\bf x})=g \delta ^{3}({\bf x})$ has been answered only recently:
In which direction does
this interaction
push the critical temperature, and by which amount? Although this problem appears to be very simple, the answer has had an
adventurous history
\cite{Baym1,Ceperley,Reppy,Wilkens1,Mueller,Holzmann1,Baym2,Stoof1,Svistunov,Arnoldnum,Ramos,Kneur1,Braaten2,kleinertbec,DELTA,Boris,Ar}.

The weak repulsion defines an interaction temperature
$T^{({\rm int})}=gn/k_B$ where $n$ is the particle density and $k_B$
the Boltzmann factor.
For small enough $g$,
the interaction temperature is
much smaller than the interaction-free critical temperature $T_c^{(0)}= 2 \pi \hbar^2 n^{2/3}/ m k_B \left[ \zeta (3/2) \right]^{2/3}$,
implying that
calculations near
$T_c^{(0)}$ can be done in the high-temperature
limit of the theory.
They can therefore
be derived from the classical limit of many-body
theory, where only the zero Matsubara modes of the Bose fields are
included \cite{Baym1}.
In this approximation, many groups calculated that the
leading shift of the critical temperature is linear in the
$s$-wave scattering length $a_s=Mg/4\pi\hbar ^2$:
\begin{eqnarray}
\label{shift9}
\frac{\Delta T_c}{T_c^{(0)}}=c_1 a_s^c{}n^{1/3} + {\cal O}\left(a_s^c{}^2n^{2/3}\right) \, .
\end{eqnarray}
The latest and most precise Monte Carlo simulations gave
a slope
$c_1=1.32\pm0.02$ \cite{Svistunov,Arnoldnum},
reasonably close to the
theoretical numbers 1.48 in Ref.~\cite{Ramos}, 1.15 in Ref.~\cite{Kneur1}, and
0.492 in Ref. \cite{Braaten2},
which were derived by resumming a
divergent perturbation series for $c_1$
using the so-called $\delta$-expansion method.
However, the application of this method to field theory is
unjustified, as pointed out in Refs. \cite{kleinertbec,DELTA},
since it does not take into account
that field theory shows anomalous dimensions
in the strong-coupling limit. This fundamental
flaw was corrected in Refs. \cite{kleinertbec,Boris}
where the field-theoretic variational perturbation theory (VPT)
developed in Refs.~\cite{PI,VPT2,Festschrift} was used to
self-consistently determine the anomalous dimension from
five- to six-loop expansions. The six-loop result
is $1.25 \pm 0.13$ \cite{Boris}, thus
agreeing with the latest Monte Carlo value
$1.32\pm0.02$ \cite{Svistunov,Arnoldnum}.

\begin{figure}[t]
\hspace*{0.1cm} \epsfxsize=10cm \epsffile{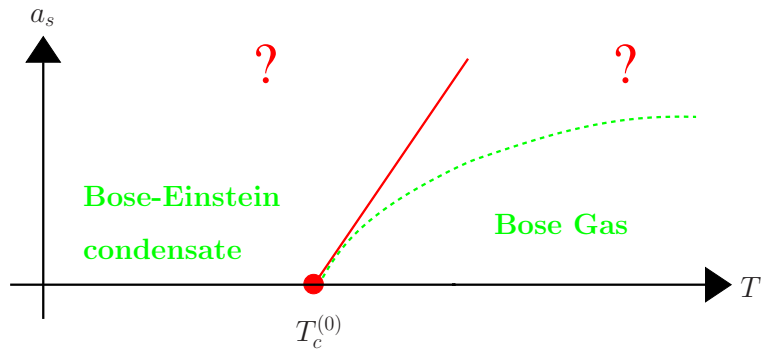}
\caption{\label{fig0} Phase diagram in the
$a_s$--$T$-plane. The straight line indicates the prediction (\ref{shift9}) of
classical field theory valid locally in the vicinity of
$T_c^{(0)}$. The dashed curve illustrates the second-order result (\ref{zeromat10}) of Arnold et al. \cite{Ar}.
The full phase diagram will be given in Fig. \ref{fig1}.}
\end{figure}

If we draw a phase diagram
in the $a_s$--$T$-plane,
the result (\ref{shift9})
gives a curve starting linearily to the right
at the critical temperature $T_c^{(0)}$, as illustrated in Fig. \ref{fig0}.
A first  correction to this curve has been
calculated recently by Arnold and co-workers \cite{Ar}
by going beyond the classical
limit. They find
\begin{eqnarray}
\label{zeromat10}
\frac{\Delta T_c}{T_c^{(0)}}=c_1 a_s^c{} n^{1/3}+a_s^c{} n^{1/3}(c_2'\ln a_s^c{} n^{1/3}+c_2) + {\cal O}\left(a_s^c{}^3 n \right)\,.
\end{eqnarray}
Their value for the coefficient
$c_2'=-64\pi\zeta(1/2)/3\zeta(3/2)^{5/3}\approx 19.75$ is determined perturbatively,
whereas $c_2$ requires again
resumming a divergent series which is unknown so far.
With the help of Monte Carlo data they estimate
$c_2\approx 75.7$. If one plots (\ref{zeromat10})
with $c_1 \approx 1.3$ in the phase diagram of Fig. \ref{fig0}, the transition
curve bends to the right. This cannot
go on for higher $a_s$ since there cannot
be a condensate for high $T$.
The next correction terms
must turn the transition curve in
Fig. \ref{fig0} to the left,
so that it arrives, for $T=0$,
at a pure quantum phase transition \cite{Sachdev}.

In this paper we follow Ref. \cite{PRL} and investigate in detail the full
quantum phase diagram. As it
is crucial to resum asymptotic perturbation series, we present
in Section \ref{vpt} a short
introduction
to VPT. In Section \ref{eff} we derive the self-consistent Popov approximation to the effective potential of a homogeneous
Bose gas using VPT.
This leads to a lowest-order
quantum phase diagram
which already shows
the interesting reentrant phenomenon.
The curve is
 variationally improved in Section \ref{bec} to include the
five- to six-loop
results near $T_c^{(0)}$
which weakens the nose in the phase diagram
but does not destroy
the reentrant transition.
\section{Variational Perturbation Theory}\label{vpt}
In this section
we briefly outline the general procedure for resumming an asymptotic
perturbation series with the help of VPT following Refs. \cite{PI,VPT2,Festschrift}.
\comment{
In order to estimate the quality of the resummation we
emphasize, in particular, how to convert a given weak-coupling
expansion into its strong-coupling limit.}
\subsection{Summary of Procedure}
\begin{figure}[t]
\hspace*{0.1cm} \epsfysize=5cm \epsffile{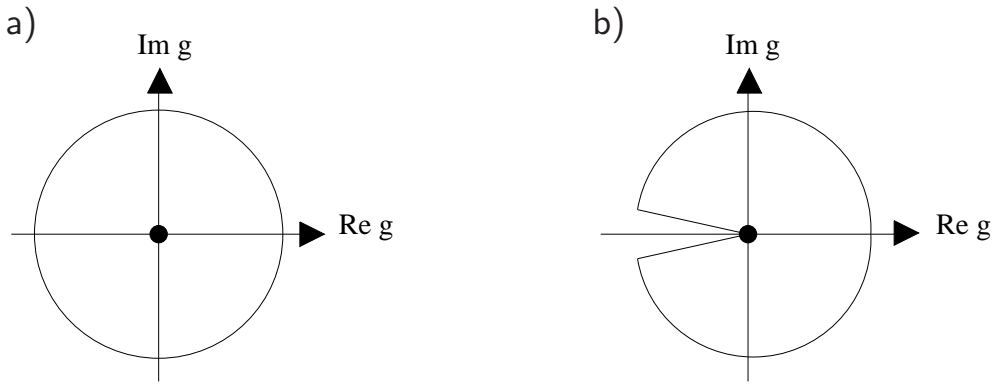}
\caption{\label{analytic} Comparing schematically analytic
properties of convergent a) and asymptotic b) series.}
\end{figure}

Perturbation expansions
can only be used to calculate
a physical quantity $f(g)$
for very small values
of the coupling constant $g$.
We are usually
able to find
a   truncated
power series at
some finite order $N$:
\begin{eqnarray}
\label{PI} f_N(g)=\sum_{n=0}^{N} a_n g^n \, .
\end{eqnarray}
If $g$ is small enough,
this can lead to an impressive
agreement between
theory and
experiment, the most prominent
example being the perturbation series for the
anomalous magnetic moment of the electron $g_{\rm e}$ where $g= \alpha \approx1/137$.
The smallness of $g$ is necessary since
the expansion (\ref{PI})
has a zero radius of convergence, as pointed out
by  Freeman Dyson in 1952 \cite{Dyson}.
The point $g=0$
is the endpoint of a cut  (see Fig. \ref{analytic})
which makes the expansion
divergent, with only an asymptotic
validity for $g\rightarrow 0$.
The difference between
a convergent and an asymptotic series
lies in the behavior
of the last term
$a_Ng^N$ of (\ref{PI}).
For fixed $g$, a convergent series has
decreasing
$a_Ng^N$, a typical divergent series of
perturbation expansions has
$a_N$ growing factorially, so that
the last term
$a_Ng^N$ decreases only at
 small $g$
for a few low orders, after
which it  increases  dramatically.

These divergent  series require resummation
if we want to extract
from them reliable
 results. The crudest method to
approximate  $f(g)$
is via Pad\'e approximants \cite{Baker}. These are rational functions
with the same power series expansions as $f(g)$.
The  Pad\'e method approximates the
left-hand cut of the function $f(g)$ in the complex $g$-plane by a
string of poles.

A better
approximation can be found by using, in addition, the knowledge of the large-order
behavior of the weak-coupling coefficients $a_n$. By means of
Borel transformations the factorial growth of $a_n$ can be
eliminated \cite{Borel,Bender}, and  a Pad\'e approximation is
applied. After this, one returns back to the original function
by an inverse Borel transform.
This procedure can be improved further by a
conformal mapping technique in which the complex $g$-plane is
mapped into a
unit circle which contains the original left-hand cut on
its circumference \cite{Map}.

A further simple tool for extracting physical results
is provided by simple variational methods
invented a long time ago by many research groups in quantum mechanics.
For instance, the
so-called $\delta$-expansion amounts to a resummation of
divergent perturbation series
(see, for instance, the references cited in \cite{DELTA}).
It is based on  introducing artificially a  harmonic trial
oscillator and  optimizing the trial frequency
by invoking the principle of minimal sensitivity \cite{Stevenson}.
It turns out that the $\delta$-expansion procedure 
corresponds to a systematic extension of a variational approach
in quantum statistics \cite{Feynman1,Feynman2,Tognetti1,Tognetti2} 
to arbitrary orders \cite{PI,Festschrift,SE} and is now called VPT.
It allows us to evaluate the
asymptotic series (\ref{PI}) for all values of the coupling
constant $g$, even in the limit $g\rightarrow \infty$.
In particular, is it possible to derive from the initial weak-coupling expansion (\ref{PI})
a strong-coupling limit which has the generic form
\begin{eqnarray}
\label{SSS} f(g) = g^{p/q} \sum_{m=0}^{\infty} b^{(m)} g^{-2m/q}
\, .
\end{eqnarray}
Here $p$ and $q$ are real growth parameters
characterizing
the strong-coupling behavior.
 For the ground-state energy of the
anharmonic oscillator with $p=1$ and $q=3$, the convergence was
shown to be exponentially fast, even
for infinite coupling strength \cite{JankeC1,JankeC2,Guida}.
It also allows us to combine a weak-coupling
expansion with information on the strong-coupling
expansion and interpolate between them \cite{Int}.

In recent years, VPT has been extended
in a simple but essential
way to become applicable
to quantum field theory
with its anomalous dimensions.
Applications to
series expansions of
$\phi^4$-theory have led to extremely accurate
critical exponents
\cite{VPT2,KleinertD,Festschrift}. The
field-theoretic
perturbation coefficients are available up to six and partly to
seven loops in $D=3$~\cite{Nickel1,Nickel2}, and up to five loops
in $D=4-\epsilon$ dimensions~\cite{Five}. The most important
feature of this new field-theoretic VPT is
that it accounts for the anomalous power approach to the
strong-coupling limit which the $\delta$-expansion cannot
accommodate.
As we see in Eq.~(\ref{SSS}), this approach is governed by an
irrational critical exponent $\omega=2/q$ as was first shown by
Wegner \cite{Wegner} in the context of critical phenomena. In
contrast to the $\delta$-expansion, the field-theoretic
variational perturbation expansions {\em cannot\/} be derived from adding
and  subtracting a harmonic term in the underlying action.
The Wegner exponent $ \omega $ is determined from (\ref{PI})
by making the variationally resummed
logarithmic derivative
$\partial \ln f(g)/\partial \ln g$ vanish at $g\rightarrow \infty$, as we have there $p=0$.
The theoretical results of the field-theoretic
VPT are in excellent agreement with experiment,
in particular
for the  only challenging
experimental value of
 the critical exponent $\alpha$ which governs the behavior of
the specific heat near the superfluid phase transition of
${}^4$He. The high experimental
accuracy is reached in one of the few
physically relevant microgravity experiments performed
in a satellite orbiting around the
earth~\cite{Satellit,Lipa}.
\subsection{Variational Expression}
Consider the truncated weak-coupling series (\ref{PI})
and rewrite it with the help
of
an auxiliary parameter $\kappa=1$ as
\begin{eqnarray}
\label{VPT2} f_N(g)=\kappa^p \sum_{n=0}^{N} a_n \left( \frac{g}
{\kappa^q} \right)^n \Big|_{\kappa=1}\,.
\end{eqnarray}
Here $p$ and $q$ denote parameters which determine the
strong-coupling behavior as we will see below. Now we introduce
a variational parameter $K$ using
 the  substitution
trick
\begin{eqnarray}
\label{VPT3}\kappa^2 = K^2({1+g r})\,,
\end{eqnarray}
where
\begin{eqnarray}
\label{VPT3b} r =  \frac{1}{g}\left(\frac{\kappa^2}{K^2}-1 \right)\,.
\end{eqnarray}
Substituting (\ref{VPT3}) into the truncated weak-coupling series
(\ref{VPT2}),
and reexpanding everything in powers of $g$ at fixed $r$,
we obtain for $ \kappa =1$:
\begin{eqnarray}
\label{VPT4} f_N(g,K)=\sum_{n=0}^N \left[ \sum_{k=0}^{N-n}
{(p-nq)/2 \choose k} \left( \frac{1}{K^2}-1 \right)^k K^{p-nq}
\right] a_n g^n \, .
\end{eqnarray}
According to the principle of minimal sensitivity
\cite{Stevenson}, we optimize the influence of $K$ on $f_N(g,K)$. At first we search
for local extrema, i.e., we determine $K$ from the condition
\begin{eqnarray}
\label{Sol1} \left. \frac{\partial f_N(g,K)}{\partial K}
\right|_{K = K_N ( g )} = 0 \, .
\end{eqnarray}
It may happen that this condition is not solvable. In this case,
we satisfy the principle of minimal sensitivity
by looking for turning points, i.e., we
fix the variational parameter according to
\begin{eqnarray}
\label{Sol2} \left. \frac{\partial^2 f_N(g,K)}{\partial K^2}
\right|_{K = K_N ( g )} = 0 \, .
\end{eqnarray}
The solutions of Eqs. (\ref{Sol1}) or (\ref{Sol2}) then yield the
variational result $f_N ( g , K_N ( g ) )$ which turns out to be a
good approximation for the function $f (g)$ for all values of the
coupling constant $g$. The quality of this approximation can be
estimated by investigating the strong-coupling limit as a special
case.
\subsection{Strong-Coupling Limit}
The conditions (\ref{Sol1}) and (\ref{Sol2})
for the function (\ref{VPT4}) imply
 that the optimal variational parameter
$K_N ( g )$ has the strong-coupling behavior
\begin{eqnarray}
\label{sahne2} K_N (g)=g^{1/q} \left( K_N^{(0)}+K_N^{(1)} g^{-2/q}
+ ... \right) \, .
\end{eqnarray}
Thus $f ( g )$ is approximated in the limit $g \rightarrow \infty$ by
\begin{eqnarray}
\label{VPT5} f_N ( g , K_N ( g ) ) = g^{p/q} \left[
b_N^{(0)}(K_N^{(0)})+b_N^{(1)}(K_N^{(0)}, K_N^{(1)}) g^{-2/q}+...
\right] \, .
\end{eqnarray}
The ratio $p/q$ determines the leading $g$-power,
 and $2/q$ the approach to scaling. The leading
strong-coupling coefficient $b_N^{(0)}(K_N^{(0)})$
is
given by
\begin{eqnarray}
\label{sahne5} b_N^{(0)}(K_N^{(0)}) =\sum_{n=0}^{N}
\sum_{k=0}^{N-n} {(p-nq)/2 \choose k}(-1)^k \,
(K_N^{(0)})^{p-nq}\,a_n ,
\end{eqnarray}
where the inner sum can be further simplified, using
the identity
\cite[see Eq.~(0.151) therein]{Gradshteyn}
\begin{eqnarray}
\label{VPT5b} \sum_{k=0}^m (-1)^k {\alpha \choose k}=(-1)^m{\alpha
-1\choose m} \, ,
\end{eqnarray}
so that the strong-coupling coefficient (\ref{sahne5})
has the variational expression
\begin{eqnarray}
\label{BL18} b_N^{(0)}(K_N^{(0)})=\sum_{n=0}^N (-1)^{N-n}
{(p-nq)/2-1 \choose N-n} (K_N^{(0)})^{p-nq} \, a_n ,
\end{eqnarray}
to be optimized in
$K_N^{(0)}$.
Inserting the optimized $K_N^{(0)}$ in (\ref{BL18}) then leads to
the variational approximation $b_N^{(0)}(K_N^{(0)})$ of the strong-coupling
coefficient $b^{(0)}$.
\section{Effective Potential}\label{eff}
In order to study
the Bose gas below the critical temperature,
we calculate the one-loop approximation to the effective potential. In this
so-called Popov approximation we determine the
number of non-condensed particles.
A variational resummation yields the self-consistent Popov approximation
which allows us to find a lowest-order expression
for the location of the quantum phase transition in the $a_s$--$T$-plane.

\subsection{One-Loop Approximation}
A grand-canonical ensemble
of Bose particles with a repulsive
two-particle $\delta$-function interaction
is governed by
the Euclidean action
\begin{eqnarray}
{\cal A}[\psi^\ast,\psi] &=& \int_0^{\hbar \beta}d\tau \int d^D x
\Big\{ \psi^\ast ({\bf x},\tau)[ \hbar\partial_\tau + \epsilon
(-i\hbar{\BF \nabla}) -\mu] \psi({\bf x},\tau)
+\frac{g}{2}\, \psi({\bf x},\tau)^2
\psi^\ast({\bf x},\tau)^2  \Big\} \,,
\end{eqnarray}
where $\epsilon({\bf k})$ denotes the one-particle energies, $\mu$
 the chemical potential,  $g
={4 \pi^2 \hbar^2 a_s}/{M} $
the coupling constant,
and $\beta\equiv 1/k_B T$. The
one-loop approximation to the
effective potential
is most easily obtained
with the help of the background method \cite[see Subsection 3.19.4 therein]{PI}.
We consider the functional integral for the partition function
\begin{eqnarray}
\label{FI}
{\cal Z} = \oint {\cal D}\psi \oint{\cal D} \psi^* \, e^{-{\cal A}[\psi,\psi^\ast]/\hbar}
\end{eqnarray}
performed
over all Bose fields $\psi({\bf x},\tau), \psi^*({\bf x},\tau)$
periodic in $\tau \in (0,\hbar  \beta )$, and
change variables to
field fluctuations $\delta \psi({\bf x},\tau), \delta \psi^*({\bf x},\tau)$  around a constant background $\Psi,\Psi^*$,
defined by
\begin{eqnarray}
\psi({\bf x},\tau)=\Psi+\delta\psi({\bf x},\tau) \, ,
\hspace*{0.5cm} \psi^*({\bf x},\tau)=\Psi^*+\delta\psi^*({\bf x},\tau) \, .
\end{eqnarray}
The functional integral (\ref{FI}) is then evaluated
including only the harmonic fluctuations $\delta\psi ({\bf x},\tau), \delta\psi^* ({\bf x},\tau)$.
The linear terms in $\delta\psi ({\bf x},\tau), \delta\psi^* ({\bf x},\tau)$
are ignored in
the background method. Thus the partition function factorizes as
${\cal Z} = {\cal Z}^{(0)} {\cal Z}^{(1)} {\cal Z}^{({\rm rest})}$, where the zero-loop term
${\cal Z}^{(0)}=e^{-{\cal A}[\Psi,\Psi^*]/\hbar}$ is given by
the tree-level
\begin{eqnarray}
{\cal A}[\Psi,\Psi^*] = \hbar\beta V \left( -\mu|\Psi|^2+\frac{g}{2}|\Psi|^4\right) \,,
\end{eqnarray}
and the one-loop contribution
\begin{eqnarray}
\label{Z1}
{\cal Z}^{(1)} = \oint {\cal D} \delta \psi \oint{\cal D} \delta \psi^* \,
e^{-{\cal A}^{({\rm quad})}[\delta \psi,\delta \psi^\ast]/\hbar}
\end{eqnarray}
involves a quadratic part whose action which can be written in matrix form:
\begin{eqnarray}
{\cal A}^{\rm (quad)}[\delta\psi,\delta\psi^\ast]
=\frac{\hbar}{2}\int_0^{\hbar\beta} d \tau \int_0^{\hbar\beta} d\tau' \int d^Dx\int d^Dx'
\left(\delta\psi^{\ast}({\bf x},\tau) ,\delta\psi({\bf x},\tau) \right)
G^{-1}({\bf x},\tau;{\bf x}',\tau')
\left( \begin{array}{c} \delta\psi({\bf x}',\tau') \\[1mm] \delta\psi^{\ast}({\bf x}',\tau')
\end{array} \right)\,.
\end{eqnarray}
The functional matrix
\begin{eqnarray}
\label{inv}
G^{-1}({\bf x},\tau;{\bf x}',\tau')=
\delta({\bf x} - {\bf x}') \delta(\tau-\tau')
\frac{1}{\hbar}
\left(\begin{array}{cc}
\hbar\partial_{\tau'}+\epsilon
(-i\hbar{\BF \nabla'})-\mu+2g|\Psi|^2 & g\Psi^2 \\
g{\Psi^{\ast}}^2 & -\hbar\partial_{\tau'}+\epsilon
(-i\hbar{\BF \nabla'})-\mu+2g|\Psi|^2
\end{array} \right) 
\end{eqnarray}
is the inverse Green function. The result of the Gaussian integral in (\ref{Z1}) reads
\begin{equation}
{\cal Z}^{(1)} = \exp \left(- \frac{1}{2} \,\mbox{Tr}\,\ln G^{-1}\right) \, .
\label{@Trace}
\end{equation}
Diagonalizing  (\ref{inv})
in momentum space
leads to
\begin{eqnarray}
\label{inv2}
G^{-1}({\bf k},\tau-\tau')= \delta(\tau-\tau') 
\frac{1}{\hbar}
\left(\begin{array}{cc}
\hbar\partial_{\tau'}+\epsilon({\bf k})- \mu + 3 g |\Psi|^2 &0 \\
0 & -\hbar\partial_{\tau'}+\epsilon({\bf k})- \mu +  g |\Psi|^2
\end{array} \right)  \, ,
\end{eqnarray}
so that (\ref{@Trace}) reduces to
\begin{eqnarray}
\label{REZ}
{\cal Z}^{(1)} = \exp \left\{- \frac{\beta}{2}  \sum_{{\bf k}}
E({\bf k}) - \sum_{{\bf k}} \ln \left[ 1-e^{-\beta E({\bf k})} \right] \right\} \, .
\end{eqnarray}
Here we have introduced the $\Psi,\Psi^*$-dependent quasiparticle energies
\begin{eqnarray}
\label{bog2}
E({\bf k})=\sqrt{[\epsilon({\bf k})-\mu+2g |\Psi|^2]^2-g^2|\Psi|^4}\,.
\end{eqnarray}
Combining the zero- and the one-loop contribution,
we obtain for the  effective potential
${\cal V} (\Psi,\Psi^\ast) \equiv  - (\ln {\cal Z})/\beta$
the simple expression
\begin{eqnarray}
\label{loop0}
{\cal V} (\Psi,\Psi^\ast) = V \left( -\mu|\Psi|^2+\frac{g}{2}|\Psi|^4 \right) +\frac{\eta}{2}
\sum_{{\bf k}}E({\bf k}) + \frac{\eta}{\beta}\sum_{{\bf k}} \ln[1-e^{-\beta E({\bf k})}]+{\cal O}(\eta^2)\,,
\label{VEFF}
\end{eqnarray}
where we have introduced an expansion parameter $\eta =1$ whose
power serves to count the loop order of each term.
\subsection{Popov Approximation}
The one-loop effective potential (\ref{loop0})
%, known as the Popov approximation \cite{Andersen,Griffin}
is still a function of the
background fields $\Psi, \Psi^*$.
If extremized in these fields,
it yields
the thermodynamic potential
at its extremum. The extremalization condition is
\begin{eqnarray}
\label{bog32}
-\mu+g|\Psi|^2+\frac{\eta}{2V}\sum_{{\bf k}}\frac{2g[\epsilon({\bf k})-\mu+2g|\Psi|^2]-g^2|\Psi|^2}{E({\bf k})}
+\frac{\eta}{V}\sum_{{\bf k}}\frac{2g[\epsilon({\bf k})-\mu+2g|\Psi|^2]-
g^2|\Psi|^2}{E({\bf k}) \left[ e^{\,\beta E({\bf k})}-1\right]}+{\cal O}(\eta^2)= 0\,.
\end{eqnarray}
The extremal value of $\Psi^* \Psi$ defines the
condensate density $n_0=|\Psi|^2$, for which
(\ref{bog32}) yields the one-loop equation
\begin{eqnarray}
\label{bog4}
n_0=\frac{\mu}{g}-\frac{\eta}{V}\sum_{{\bf k}}\frac{2\epsilon({\bf k})+\mu}{\sqrt{\epsilon({\bf k})^2+2\mu
\epsilon({\bf k})}}\left(\frac{1}{2}+\frac{1}{e^{\,\beta
\sqrt{\epsilon({\bf k})^2+2\mu \epsilon({\bf k})}}-1}\right)+{\cal O}(\eta^2)\,.
\end{eqnarray}
For $\eta =0$ we obtain the tree-level condensate density $n_0^{(0)}=\mu/g$, 
which leads in (\ref{bog2}) to the quasiparticle energies
found first  by Bogoliubov \cite{Bogoliubov}:
\begin{eqnarray}
\label{bog5}
E^{(0)}({\bf k})=\sqrt{\epsilon({\bf k})^2+2\mu \epsilon({\bf k})}\,.
\end{eqnarray}
Inserting (\ref{bog4}) into (\ref{loop0}) yields the grand-canonical potential:
\begin{eqnarray}
\label{bog6}
\frac{{\Omega}(\mu,T)}{V}=-\frac{\mu^2}{2g}+\frac{\eta}{2V}\sum_{{\bf k}}\sqrt{\epsilon({\bf k})^2+2\mu \epsilon({\bf k})}+
\frac{\eta}{\beta V}\sum_{{\bf k}}\ln \left(1-e^{-\beta \sqrt{\epsilon({\bf k})^2+2\mu \epsilon({\bf k})}}\,\,\right)
+{\cal O}(\eta^2)\,.
\end{eqnarray}
From this, we obtain the total particle density  from the derivative
\begin{eqnarray}
\label{n}
n(\mu,T)=-\frac{1}{V} \left. \frac{\partial \Omega(\mu,T)}{\partial\mu} \right|_{T}
\end{eqnarray}
which has the explicit form
\begin{eqnarray}
\label{bog7}
n=\frac{\mu}{g}-\frac{\eta}{V}\sum_{{\bf k}}\frac{\epsilon({\bf k})}{\sqrt{\epsilon({\bf k})^2+2\mu
\epsilon({\bf k})}}\left(\frac{1}{2}+\frac{1}{e^{\,\beta
\sqrt{\epsilon({\bf k})^2+2\mu \epsilon({\bf k})}}-1}\right)+\mathcal{O}(\eta^2)\,.
\end{eqnarray}
Since the interesting experimental quantity
is the condensate density $n_0$, we eliminate
the chemical potantial $\mu$ in favor of $n_0$ via
(\ref{bog4}) and find
\begin{eqnarray}
\label{bog8}
n-n_0=\frac{\eta}{V}\sum_{{\bf k}}\frac{\epsilon({\bf k})
+gn_0}{\sqrt{\epsilon({\bf k})^2+2gn_0 \epsilon({\bf k})}}
\left(\frac{1}{2}+\frac{1}{e^{\,\beta \sqrt{\epsilon({\bf k})^2+2gn_0 \epsilon({\bf k})}}-1}\right)+\mathcal{O}(\eta^2)\,.
\end{eqnarray}
The right-hand side of Eq. (\ref{bog8}) denotes the number of
non-condensed particles in the
so-called  Popov approximation
\cite{Andersen,Griffin}.
\subsection{Self-Consistent Popov Approximation}\label{popovapprox}
The result (\ref{bog8}) is valid only for
$n\approx n_0$, i.e. for small $\eta$. The
quantum phase transition \cite{Sachdev} which we want to
locate, however,
takes place for $n\gg n_0$, i.e. at the large value $\eta =1$
of the loop counter. This strong-coupling regime
can be reached by applying VPT, following the rules
reviewed in Section \ref{vpt}. Here
we introduce a variational parameter $M$ by replacing,
as in (\ref{VPT3}) and (\ref{VPT3b}),
\begin{eqnarray}
\label{pop1}
\mu= M+r\eta,
\end{eqnarray}
with the abbreviation
\begin{eqnarray}
\label{pop2}
r=\frac{\mu-M}{\eta}\,.
\end{eqnarray}
Inserting (\ref{pop1}) into the grand-canonical potential
(\ref{bog6}), and reexpanding this
in powers of $ \eta $ at fixed $r$, we obtain
\begin{eqnarray}
\label{pop3}
\frac{\Omega^{\rm trial}(M,\mu,T)}{V}=-\frac{M^2}{2g}-\eta\frac{M}{g}r+\frac{\eta}{2V}
\sum_{{\bf k}}\sqrt{\epsilon({\bf k})^2+2M \epsilon({\bf k})}
+\frac{\eta}{\beta V}\sum_{{\bf k}}\ln\left(1-e^{-\beta
\sqrt{\epsilon({\bf k})^2+2M \epsilon({\bf k})}}\,\right) +
\mathcal{O}(\eta^2) \, .
\end{eqnarray}
Reinserting back $r$ from (\ref{pop2}), ignoring the terms
$\mathcal{O}(\eta^2)$, and setting $\eta =1$,
we arrive at the following trial function for the
grand-canonical potential:
\begin{eqnarray}
\label{pop4}
\frac{\Omega^{\rm trial}(M,\mu,T)}{V}=\frac{M^2}{2g}-\frac{M\mu}{g}+
\frac{1}{2V}\sum_{{\bf k}}\sqrt{\epsilon({\bf k})^2+2M \epsilon({\bf k})}
+ \frac{1}{\beta V}\sum_{{\bf k}}\ln\left(1-e^{-\beta
\sqrt{\epsilon({\bf k})^2+2M \epsilon({\bf k})}}\,\right).
\end{eqnarray}
This is optimized
with respect to the variational parameter
$M$, yielding
\begin{eqnarray}
\label{pop6}
M^{\rm opt} = \mu-\frac{g}{V}\sum_{{\bf k}}
\frac{\epsilon({\bf k})}{\sqrt{\epsilon({\bf k})^2+2M^{\rm opt}
\epsilon({\bf k})}}\left(\frac{1}{2}+\frac{1}{e^{\,\beta
\sqrt{\epsilon({\bf k})^2+2M^{\rm opt}\epsilon({\bf k})}}-1}\right)
.
\end{eqnarray}
Inserting (\ref{pop6}) into (\ref{pop4}) yields the optimized
grand-canonical potential
\begin{eqnarray}
\label{pop7}
{\Omega}(\mu,T)={\Omega}^{\rm trial}(M^{\rm opt},\mu,T)\,.
\end{eqnarray}
However, we are more interested in a resummation of equation
(\ref{bog8}). To this end we compute the particle density (\ref{n}) by using (\ref{pop7})
\begin{eqnarray}
\label{pop8}
n=-\left.\frac{1}{V}\frac{\partial{\Omega}^{\rm trial}(M^{\rm opt},\mu,T)}{\partial\mu}\right|_{M^{\rm opt},T}
-\frac{1}{V}\left.\frac{\partial{\Omega}^{\rm trial}(M^{\rm opt},\mu,T)}{\partial M^{\rm opt}}\right|_{\mu,T}
\frac{\partial M^{\rm opt}}{\partial\mu}\,.
\end{eqnarray}
Because of Eqs. (\ref{pop4}) and (\ref{pop6}), this reduces to
\begin{eqnarray}
\label{pop9}
n=\frac{M^{\rm opt}}{g}\,.
\end{eqnarray}
Furthermore, we perform a similar variational resummation 
for the condensate
density (\ref{bog4}).
The substitution
(\ref{pop1}) yields the variational expression
\begin{eqnarray}
\label{pop10}
n_0=\frac{M}{g}+\eta\frac{r}{g}-\frac{\eta}{V}\sum_{{\bf k}}
\frac{2\epsilon({\bf k})+M}{\sqrt{\epsilon({\bf k})^2+2M
\epsilon({\bf k})}}\left(\frac{1}{2}+\frac{1}{e^{\,\beta
\sqrt{\epsilon({\bf k})^2+2M\epsilon({\bf k})}}-1}\right)+ \mathcal{O}(\eta^2) \,.
\end{eqnarray}
Reinserting (\ref{pop2}) and ignoring the terms $\mathcal{O}(\eta^2)$, this becomes  at $\eta =1$
\begin{eqnarray}
\label{pop11}
n_0=\frac{\mu}{g}-\frac{1}{V}\sum_{{\bf k}}\frac{2\epsilon({\bf k})+M}{\sqrt{\epsilon({\bf k})^2+2M
\epsilon({\bf k})}}\left(\frac{1}{2}+\frac{1}{e^{\,\beta
\sqrt{\epsilon({\bf k})^2+2M\epsilon({\bf k})}}-1}\right)\,.
\end{eqnarray}
Evaluating this expression
at $M=M^{\rm opt}$ with (\ref{pop6}),
we find by taking into account (\ref{pop9})
\begin{eqnarray}
\label{bog8b}
n-n_0=\frac{1}{V}\sum_{{\bf k}}\frac{\epsilon({\bf k})
+gn}{\sqrt{\epsilon({\bf k})^2+2gn \epsilon({\bf k})}}
\left(\frac{1}{2}+\frac{1}{e^{\,\beta \sqrt{\epsilon({\bf k})^2+2gn \epsilon({\bf k})}}-1}\right)\,.
\end{eqnarray}
Thus we have arrived at
the self-consistent Popov approximation, which differs from the
usual Popov approximation (\ref{bog8})
by having
the total particle  density $n$ on the right-hand side
rather than the
condensate
density $n_0$.
Note that Eq. (\ref{bog8b}) could also be obtained by just replacing $n_0$ in the right-hand side of Eq. (\ref{bog8})
by $n$, the error involved in this substitution being of order ${\cal O} (\eta^2)$.
The above VPT analysis puts such a ``dressing'' procedure on a rigorous footing.
Furthermore, we remark that
the location of the quantum phase transition \cite{Sachdev} for all $T$ is obtained from (\ref{bog8b})
by solving this equation for $n_0=0$ \cite{Bogoliubov,Abrikosov}.
\section{Lowest-Order Transition Curve}\label{bec}
In this section we determine the quantum phase transitions
for a homogeneous BEC
in the $a_s$--$T$-plane
from (\ref{bog8b}).
\subsection{Zero-Temperature Limit}
Inserting into (\ref{bog8b}) the free-particle energies
\begin{eqnarray}
\label{spectrum1}
\epsilon({\bf k})=\frac{\hbar^2{\bf k}^2}{2m}\,,
\end{eqnarray}
and going to the
thermodynamic limit, in which we can replace
the momentum sum by an integral
\begin{eqnarray}
\label{trafo1}
\sum_{{\bf k}}\rightarrow V \int \frac{d^D k}{(2\pi)^D}\,,
\end{eqnarray}
we find the zero-temperature equation
\begin{eqnarray}
\label{zero2}
n = n_0+\frac{1}{2}
\int
\frac{d^D k}{(2\pi)^D}
\frac{1}{E({\bf k})}
\left(
\frac{\hbar^2{\bf k}^2}
{2m}+gn
\right),
\end{eqnarray}
with the
Bogoliubov quasi-particle
energies
\begin{eqnarray}
\label{BS}
E({\bf k}) = \sqrt{\left(\frac{\hbar^2{\bf k}^2}{2m}\right)^2
+2gn\,\frac{\hbar^2{\bf k}^2}{2m} }\, .
\end{eqnarray}
For large values of ${\bf k}$, the Bogoliubov energies
coincide with (\ref{spectrum1}).
For small values of ${\bf k}$, however, the Bogoliubov spectrum (\ref{BS}) has a linear
${\bf k}$-dependence
$E({\bf k}) \approx c \, \hbar |{\bf k}|$, where $c$
is  the velocity of second sound:
\begin{eqnarray}
\label{zero21}
c=\sqrt{\frac{4\pi\hbar^2na_s}{m^2}}.
\end{eqnarray}
Since $E({\bf k})/|{\bf k}|$
is bounded from below, the interacting Bose gas is superfluid.
Recently, the Bogoliubov sound velocity (\ref{zero21})
was measured in a trapped BEC at MIT \cite{Ketterle}.

As the integrand in (\ref{zero2}) depends only on the
absolute value of the wave vector ${\bf k}$,
the integrals are effectively one-dimensional,
and we may rewrite (\ref{zero2}) as
\begin{eqnarray}
\label{zero4}
n-n_0=\frac{1}{2}\left[J_D \left(1,\frac{1}{2},2gn\right)+gn\,J_D \left(0,\frac{1}{2},2gn\right)\right]\,.
\end{eqnarray}
Here we have introduced the master integral
\begin{eqnarray}
\label{zero5}
J_D (\alpha,\beta,a)=\frac{1}{\Gamma(D/2)}\left(\frac{m}{2\pi\hbar^2}\right)^{D/2}
\int_0^\infty dx \, \frac{x^{\alpha+D/2-\beta-1}}{(x + a)^\beta} \, ,
\end{eqnarray}
which is related to the integral representation of the Beta function
\begin{equation}
B(x,y) \equiv \frac{\Gamma (x)  \Gamma (y)}{\Gamma (x+y)} =\int_0^\infty
dt\, \frac{t^{x-1}}{(t+1)^{x+y}}
\label{@Beta}
\end{equation}
according to
\begin{eqnarray}
J_D (\alpha,\beta,a)= \frac{a^{D/2+\alpha-2\beta}}{\Gamma(D/2)}\left(\frac{m}{2\pi\hbar^2}\right)^{D/2}
B\left(\frac{D}{2}+\alpha-\beta, -\frac{D}{2}-\alpha+2\beta\right)
\,.
\end{eqnarray}
Thus we find for the particle density (\ref{zero4}) in $D=3$ dimensions
the equation:
\begin{eqnarray}
\label{zero12} \frac{n-n_0}{n}=
\frac{ \sqrt{ m^{3}g^{3}n}}{3\pi^2 \hbar^3}\,,
\end{eqnarray}
which gives the so-called {\em depletion\/}
of the condensate.
Expressing $g$ in terms
of
the
$s$-wave scattering length $a_s$ leads to the
well-known Bogoliubov depletion \cite{Bogoliubov}:
\begin{eqnarray}
\label{zero13}
\frac{n-n_0}{n}=\frac{8}{3}\sqrt{\frac{a_s^3n}{\pi}} \,.
\end{eqnarray}
Since this self-consistent equation
was derived by variational perturbation theory
which we know to be very good up to strong couplings \cite{PI},
we may use it to find the
strong-coupling value
of $a_s$ where the depletion
is complete.
This is the point of a quantum phase transition
\cite{Sachdev}
at zero temperature, taking
place at 
\begin{equation}
a_s^c{} (T=0) n^{1/3}= \left(\frac{9\pi}{64}\right)^{1/3} \, .
\label{@}
\end{equation}
\subsection{Full Quantum Phase Diagram}
We now calulate the temperature dependence
of the quantum phase transition. Using (\ref{bog8b})--(\ref{trafo1}), we obtain instead of
the zero-temperature equation (\ref{zero12})
\begin{eqnarray}
\label{nonzero1}
a_s\,n^{1/3}\left[1+\frac{3\alpha}{16}I(\alpha)\right]^{2/3}
=\left(\frac{n-n_0}{n}\right)^{2/3}\left(\frac{9\pi}{64}\right)^{1/3} \,,
\end{eqnarray}
where $I( \alpha )$ abbreviates the integral
\begin{eqnarray}
\label{nonzero2} I(\alpha)=\int_0^\infty dx \,
\frac{x\alpha+8}{2\sqrt{x\alpha +16}
\,\, ( e^{\, \sqrt{x^2\alpha/16 +x}}-1) }\,.
\end{eqnarray}
The
dimensionless parameter
$\alpha$ is given by
\begin{eqnarray}
\label{nonzero3}
\alpha=\left[\frac{t}{a_sn^{1/3}\zeta(3/2)^{2/3}}\right]^2\,,
\end{eqnarray}
where $t=T/T_c^{(0)}$ denotes the reduced temperature.
 The quantum phase transition
at
$n_0=0$ obeys now
the equation
\begin{eqnarray}
\label{phase1}
a_s^c{}\,n^{1/3}\left[1+\frac{3\alpha_c}{16}I(\alpha_c)\right]^{2/3}  =\left(\frac{9\pi}{64}\right)^{1/3} \,.
\end{eqnarray}
Near $T=0$ we perform a Taylor expansion in the parameter $\alpha_c$, i.e.
\begin{eqnarray}
\label{phase2}
a_s^c{}n^{1/3}=\sum_{k=0}^N a_{k}\alpha_c^k + {\cal O}(\alpha_c^{N+1}) \, ,
\end{eqnarray}
and obtain for the coefficients:
\begin{eqnarray}
\label{phase22}
a_0=\left(\frac{9\pi}{64}\right)^{1/3}\approx 0.762\,, \hspace*{0.5cm}
a_1\approx-0.313\,,\hspace*{0.5cm}
a_2\approx0.200 \,,\hspace*{0.5cm}
a_3\approx-0.207\,.
\end{eqnarray}
The
series
has Borel character ---
the sign of the coefficients is alternating.
This  expansion is inapplicable for small
$a_s$
since
$\alpha$ diverges
in this limit.
There we  solve (\ref{phase1})
\begin{figure}[t]
\hspace*{0.1cm} \epsfxsize=10cm \epsffile{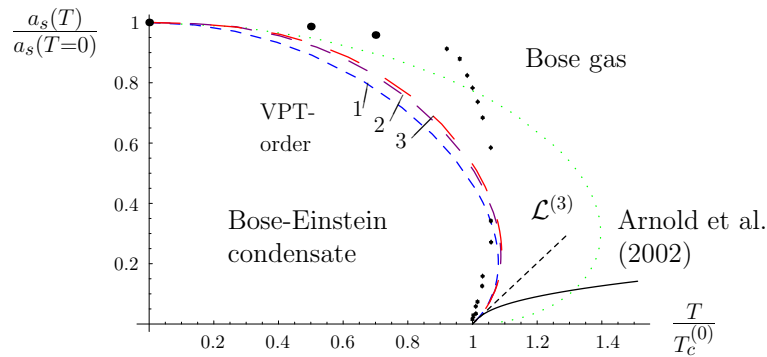}
\caption{\label{fig1}
Quantum phase diagram
of Bose-Einstein condensation in
variationally improved
one-loop approximation without (dotted)
and with properly imposed  higher-loop slope properties
at $T_c^{(0)}$ (dashed length increasing with VPT order).
Short solid curve starting at $T_c^{(0)}$ is due to Arnold et al. \cite{Ar}.
Dashed straight line indicates the slope of our curve extracted either from
Monte-Carlo data \cite{Svistunov,Arnoldnum} or recent analytic
results \cite{kleinertbec,Boris}.
Diamonds correspond to the Monte-Carlo data of Ref.
\cite{Ceperley} and dots stem from Ref. \cite{helium},
both scaled to their critical value $a_s(T=0)\approx 0.63$.}
\end{figure}
by rescaling
and expanding the integrand in (\ref{nonzero2}), yielding
\begin{eqnarray}
\label{shift1} I(\alpha)&=&\frac{\sqrt{\alpha}}{2}\int_0^\infty dz
\sqrt{z}\frac{1}{{\rm exp}\left(\,\sqrt{\alpha}z/4+2/\sqrt{\alpha}\right)-1}+\mathcal{O}\left(\frac{1}{\alpha}\right)\,.
\end{eqnarray}
Using the series representation of the Bose distribution function, we obtain
\begin{eqnarray}
\label{shift4}
I(\alpha)=\frac{4\Gamma(3/2)}{\alpha^{1/4}}\,\zeta_{3/2}(e^{-2/\sqrt{\alpha}})+{\cal O}\left(\frac{1}{\alpha}\right)\, ,
\end{eqnarray}
where we have introduced the polylogarithmic function
\begin{eqnarray}
\zeta_{\nu} ( z ) = \sum_{n=1}^{\infty} \frac{z^n}{n^{\nu}} \, .
\end{eqnarray}
It has the Robinson expansion \cite{Robinson}
\begin{eqnarray}
\label{rob4}
\zeta_\nu (e^{\,\beta\mu}) = \Gamma(1-\nu)(-\beta\mu)^{\nu-1} + \sum_{k=0}^{\infty}
\frac{(\beta\mu)^k}{k!}\,\zeta(\nu-k),\hspace*{0.5cm} \mu<0 \,,
\end{eqnarray}
which can easily be proved
with the help of
Poisson's summation  formula \cite[Chap. 2]{PI}.
Thus we obtain  for large $\alpha$:
\begin{eqnarray}
\label{shift5}
I(\alpha)=\frac{2\sqrt{\pi}\zeta(3/2)}{\alpha^{1/4}}-
\frac{4\sqrt{2}\pi}{\sqrt{\alpha}}+\mathcal{O}\left(\frac{1}{\alpha^{3/4}}\right)\,.
\end{eqnarray}
Inserting this into (\ref{phase1}) yields:
\begin{eqnarray}
\label{shift6}
1=t_c^{3/2}-\frac{2\sqrt{2\pi a_s^c{}n^{1/3}}t_c}{\zeta(3/2)^{2/3}} \,.
\end{eqnarray}
From this equation we can derive
a shift in the critical temperature due to the interaction. We expand
\begin{eqnarray}
\label{shift7} t_c=1+\frac{\Delta T_c}{T_c^{(0)}}\,,
\end{eqnarray}
and obtain from (\ref{shift6})
\begin{eqnarray}
\label{shift8}
\frac{\Delta T_c}{T_c^{(0)}}=\frac{4\sqrt{2\pi}}{3\zeta(3/2)^{2/3}}\sqrt{a_s^c{}n^{1/3}}+\mathcal{O}(a_s^c{}n^{1/3})\,.
\end{eqnarray}
This has the square-root behavior found before in Refs. \cite{Toyoda,Huang}, with the positive sign agreeing with Ref. \cite{Huang}.
The full phase diagram is displayed in
Fig. \ref{fig1}.
It shows
the interesting phenomenon of  a {\em reentrant
transition\/} above the critical temperature
$T_c^{(0)}$ of the free system, which shows up as a {\em nose\/} in the
transition curve.
This implies that a condensate can be produced
above
$T_c^{(0)}$ by {\em
increasing\/} $a_s$, which disappears upon
a further increase of $a_s$.
The transition curve differs so far
considerable
from
early
Monte-Carlo simulations \cite{Ceperley}, also
shown in Fig. \ref{fig1}.

From recent work we know,
however, that the square-root behavior (\ref{shift8}) is incorrect, since its derivation
ignores the pile-up of infrared singularities
to high orders
at the critical point.
As explained in the beginning,
Monte-Carlo simulations \cite{Svistunov,Arnoldnum}
and precise high-temperature calculations \cite{kleinertbec,Boris}
have derived  a
temperature shift
(\ref{shift9}) which is
linear in the scattering length $a_s$,
with
a slope $c_1\approx1.3$. Thus it becomes
necessary to improve our resummed one-loop approximation
(\ref{phase1}) for the transition line via VPT
in such a way that the
$\alpha_c\rightarrow\infty$ behavior of the expansion
(\ref{phase2}) is given by (\ref{shift9}) and not by (\ref{shift8}).
In doing so we can follow the interpolation program
between weak and strong-coupling
expansions
which was developed
in Ref.~\cite{Int}.
\subsection{Variationally Improved Quantum Phase Diagram}
We now calculate a more reliable transition curve for the quantum phase diagram
interpolating the weak-coupling expansions (\ref{phase2})
to a strong-coupling expansion which is compatible with
the linear increase (\ref{shift9}). The generic
strong-coupling
expansion is
\begin{eqnarray}
\label{impro2}
a_s^c{}n^{1/3}=\alpha_c^{p/q}\sum_{k=0}^N b_{k}\alpha_c^{-2k/q}.
\end{eqnarray}
We shall use only the accurate knowledge
of the small-$a_s$ behavior
(\ref{shift9}) so that
we possess only the two leading terms
in (\ref{impro2}):
\begin{eqnarray}
\label{impro4}
a_s^c{}n^{1/3}=\alpha_c^{p/q}\left(b_0+b_1\alpha_c^{-2/q}\right)\,.
\end{eqnarray}
Expressed in terms of the parameter $\alpha$ of Eq.~(\ref{nonzero3})
and replacing $t_c$ via (\ref{shift7}),
this reads:
\begin{eqnarray}
\label{impro41}
b_0-\left(a_s^c{}n^{1/3}\right)^{1+2p/q}\zeta(3/2)^{4p/3q}+b_0\,\frac{2p}{q}\frac{\Delta
T_c}{T_c^{(0)}}+b_1\zeta(3/2)^{8/3q}\left(a_s^c{}n^{1/3}\right)^{4/q}=0\,.
\end{eqnarray}
This equation is satisfied if
\begin{eqnarray}
1+\frac{2p}{q}=0\,,\qquad \frac{4}{q}=1\,,
\end{eqnarray}
which fixes the values of $p$, $q$ to
$p=-2$, $q=4$.
Comparison between (\ref{shift9}) and (\ref{impro41}) yields
\begin{eqnarray}
\label{impro7}
b_0=\frac{1}{\zeta(3/2)^{2/3}}\,,\qquad\,b_1=\frac{c_1}{\zeta(3/2)^{4/3}}\,.
\end{eqnarray}
This result will now
be combined with our
self-consistent
one-loop approximation
(\ref{phase1}) for small $ \alpha $.
The philosophy will be to
use only the first $W+1$ weak-coupling coefficients $a_0,\dots, a_W$
 in (\ref{phase2}). We allow for two extra
weak-coupling coefficients $\tilde{a}_{W+1}$ and $\tilde{a}_{W+2}$
chosen
such that the strong-coupling
limit calculated with VPT
matches
the
coefficients $b_0$ and $b_1$ in (\ref{impro7}).
Afterwards we determine the whole quantum phase diagram
from the variationally  resummed series
\begin{eqnarray}
\label{impro8}
a_s^c{}n^{1/3}=\sum_{k=0}^{W}
a_k\alpha_c^k+\tilde{a}_{W+1}\alpha_c^{W+1}+\tilde{a}_{W+2}\alpha_c^{W+2}\,.
\end{eqnarray}

Let us illustrate this procedure for $W=0$ and determine the
subsequent weak-coupling coefficients $\tilde{a}_1,\tilde{a}_2$
from the strong-coupling coefficients $b_0,b_1$. To this end we
follow the VPT procedure of
Section \ref{vpt} by identifying $\alpha_c\equiv g$ and by
specifying (\ref{VPT4}) for $N=2$, $p=-2$, and $q=4$:
\begin{eqnarray}
\label{impro9}
f_2(g,K)=\frac{3a_0}{K^2}-\frac{3a_0}{K^4}+\frac{a_0+4g\tilde{a}_1}{K^6}-\frac{3g\tilde{a}_1}{K^8}
+\frac{\tilde{a}_2g^2}{K^{10}} \, ,
\end{eqnarray}
where the last two coefficients in (\ref{VPT4}) were replaced by
$\tilde{a}_1,\tilde{a}_2$. According to the principle of minimal
sensitivity \cite{Stevenson} we have to optimize (\ref{impro9}) with respect to $K$ and solve Eq. (\ref{Sol1})
which leads to
\begin{eqnarray}
\label{EQQ}
-\frac{6a_0}{K^3}+\frac{12a_0}{K^5}-6\frac{a_0+4g\tilde{a}_1}{K^7}+\frac{24g\tilde{a}_1}{K^9}
-\frac{10\tilde{a}_2g^2}{K^{11}} = 0 \, .
\end{eqnarray}
Inserting the ansatz (\ref{sahne2}) with
the two leading coefficients $K_2^{(0)}$ and $K_2^{(1)}$, i.e.
\begin{eqnarray}
\label{impro10}
K_2(g)=K_2^{(0)}g^{1/4}+K_2^{(1)}g^{-1/4}+\mathcal{O}(g^{-3/4})\,,
\end{eqnarray}
into (\ref{EQQ}) and comparing the coefficients of the two leading powers
$g^{-1/2}\,,\,g^{-1/4}$ in the coupling constant $g$, yields the
following two equations:
\begin{eqnarray}
\label{impro11}
0&=&3{K_2^{(0)}}^8a_0+12{K_2^{(1)}}^4\tilde{a}_1+5\tilde{a}_2\,,\\
\label{impro12}
0&=&6{K_2^{(0)}}^7a_0+9{K_2^{(0)}}^8{K_2^{(1)}}a_0+12{K_2^{(0)}}^3\tilde{a}_1
+84{K_2^{(0)}}^4{K_2^{(1)}}\tilde{a}_1+55{K_2^{(1)}}\tilde{a}_2\,.
\end{eqnarray}
As we have four unknown variables
$\tilde{a}_1,\tilde{a}_2,K_2^{(0)}$ and $K_2^{(1)}$, we need two
more equations for their unique determination. It turns out that they can be
obtained from the known strong-coupling coefficients $b_0$ and
$b_1$ in (\ref{impro7}) as follows. Inserting (\ref{impro10}) into
(\ref{impro9}) and comparing with (\ref{VPT5}) yields:
\begin{eqnarray}
\label{impro13}
b_0&=&\frac{1}{{K_2^{(0)}}^{10}}\left(3{K_2^{(0)}}^8a_0+4{K_2^{(0)}}^4\tilde{a}_1+\tilde{a}_2\right)\,,\\
\label{impro14}
b_1&=&\frac{1}{{K_2^{(0)}}^{11}}\bigg(-3{K_2^{(0)}}^7a_0+6{K_2^{(0)}}^8K_2^{(1)}a_0
+3{K_2^{(0)}}^3\tilde{a}_1
+24{K_2^{(0)}}^4K_2^{(1)}\tilde{a}_1+10K_2^{(1)}\tilde{a}_2\bigg)\, .
\end{eqnarray}
From (\ref{impro11}) and (\ref{impro13}), we determine $\tilde{a}_1$ and
$\tilde{a}_2$ as functions of $K_2^{(0)}$:
\begin{eqnarray}
\label{impro15}
\tilde{a}_1=-\frac{3}{2}\,{K_2^{(0)}}^4\left(a_0-\frac{5}{12}\,b_0{K_2^{(0)}}^2\right)\,,\hspace*{1cm}
\tilde{a}_2=3{K_2^{(0)}}^8\left(a_0-\frac{1}{2}\,b_0{K_2^{(0)}}^2\right)\,.
\end{eqnarray}
Furthermore, Eq. (\ref{impro12}) can be solved for $K_2^{(1)}$:
\begin{eqnarray}
\label{impro17}
K_2^{(1)}=-{K_2^{(0)}}^3\,\frac{b_1{K_2^{(0)}}^8+3a_0{K_2^{(0)}}^4+3\tilde{a}_1}{6{K_2^{(0)}}^8a_0+24{K_2^{(0)}}^4\tilde{a}_1+10
\tilde{a}_2}\,.
\end{eqnarray}
This expression has the interesting property, that the denominator
is zero, if we insert (\ref{impro15}). As
$K_2^{(1)}$ should be a finite quantity, we have to demand that
the numerator also vanishes. This leads to an explicit algebraic
expression for $K_2^{(0)}$, which is solved by
\begin{eqnarray}
\label{impro18}
K_2^{(0)}=\pm\left(-\frac{3a_0}{2b_1}+\sqrt{\frac{9a_0^2}{4b_1^2}-\frac{3a_1}{b_1}}\,\,\right)^{1/4}\,.
\end{eqnarray}
Note that with that the choice (\ref{impro14}) is satisfied although
it was not needed for deriving (\ref{impro18}). We
now insert the weak-coupling coefficient $a_0\approx 0.762$ from
(\ref{phase22}) and the correct strong-coupling coefficients
(\ref{impro7}) with $c_1\approx 1.3$ to yield $K_2^{(0)}\approx
\pm 0.93461$. This result leads via (\ref{impro15}) to the new coefficients:
\begin{eqnarray}
\label{impro19}
W=0: \qquad \tilde{a}_1\approx-0.654\,,\quad\qquad\tilde{a}_2\approx0.935\,.
\end{eqnarray}
Finally, the trial function (\ref{impro9}) follows to be
\begin{eqnarray}
\label{impro20}
f_2(g,K)&\approx&2.284\frac{1}{K^2}-2.284\frac{1}{K^4}+(0.761-2.616g)\frac{1}{K^6}+1.962g\frac{1}{K^8} +0.935g^2\frac{1}{K^{10}}
\end{eqnarray}
and leads together with the optimization (\ref{Sol1}) to our first resummation
improved transition line which is valid for arbitrary values of the
coupling constant. The result, shown in Fig. \ref{fig1} as VPT
order 1, has now the correct asymptotic behavior near $T_c^{(0)}$
as well as near $T=0$. In a similar way we also determined
the new computed weak-coupling coefficients of
the improved resummed one-loop approximation for the two subsequent orders
$W=1,2$:
\begin{eqnarray}
\label{impro211}
W&=&1:\qquad\tilde{a}_2=-1.864\, , \qquad\tilde{a}_3=16.66\, ,\\
\label{impro212}
W&=&2:\qquad\tilde{a}_3=-29.53\, , \qquad\tilde{a}_4=622.0\, .
\end{eqnarray}
A resummation of the corresponding weak-coupling series
(\ref{impro8}) shows the fast converging phase curves with VPT
order 2 and 3
in Fig. \ref{fig1}.
It is interesting that the new
computed coefficients in (\ref{impro211}) and (\ref{impro212})
deviate significantly from the original ones in (\ref{phase22}).
The reason is that the influence of higher orders becomes smaller
in a weak-coupling expansion and it needs higher deviations to
obtain the correct strong-coupling behavior.

The agreement with the Monte Carlo points
in Fig. \ref{fig1} is now much better than with the
initial one-loop transition curve.
However, it is not perfect since the simulations
are quite inaccurate  in the nose region.
In fact, the
slope
parameter of the Monte Carlo data is
$c_1\approx 0.3$ \cite{Ceperley}, which is a factor 4 smaller than
the true value.
The disagreement is therefore no cause of worry.

Let us end by remarking
that a nose in the phase diagram was also
found in Ref. \cite{Denteneer}. The authors did not, however,
take this phenomenon
seriously but
considered it as an artefact of their slave boson approach.

\section{Summary and Outlook}
In this article we have investigated the condensation of
a homogeneous Bose gas in the $a_s$--$T$-plane.
The result was obtained by
applying VPT at two stages.
First, we variationally resummed the Popov approximation to the effective potential in order to reach the quantum phase
transition. Second, the lowest-order
 quantum phase diagram was variationally improved
by taking into account recent results on the leading
temperature shift for small scattering length $a_s$.

Finally, we mention that the self-consistent Popov approximation is also applicable to BECs trapped in optical lattices \cite{PRL}.
There the periodicity of the lattice leads to a quasi-free behavior due to Bloch's theory which enables us to treat this system
as effectively homogeneous \cite{Fisher,Sheshadri,Freericks,Jaksch,Stoof,Zwerger}.
At $T=0$, this yields
a quantum phase transition from a superfluid phase to a
Mott insulator, in good agreement
 with recent experimental data \cite{Greiner}. The
theory
in Ref. \cite{PRL} predicts again
a reentrant phenomenon in the quantum phase diagram which might be
observed
experimentally.
The only problem is the need of a soft
magnetic trap
with frequency of the order of 10 Hz which holds the atoms
in the optical lattice.
This could in principle
destroy
the reentrant transition, since the temperature shift
in a trap is negative. It is not yet clear whether for very
small trap frequency, there exists a crossover from negative
to positive slope $c_1$.

\section*{Acknowledgements}
We thank Robert Graham, Franck Lalo{\"e},
and Flavio Nogueira for clarifying discussions as well as
the Priority Program SPP 1116 {\it Interactions in Ultra-Cold Atomic and Molecular Gases}
of the German Research Foundation (DFG) for
financial support.
The research
was also
partially supported by
the ESF COSLAB Program and by the
DFG (Deutsche Forschungsgemeinschaft)
under Grant Kl-256.
One of us (S.S.)
acknowledges support from the German National Academic Foundation.
\end{document}